\documentclass[journal]{IEEEtran}
\usepackage{amsmath,amsfonts}
\usepackage{amssymb}
\usepackage{algpseudocode}
\usepackage{bbm}
\usepackage{algorithm}
\usepackage{array}
\usepackage{color}
\usepackage{caption}
\usepackage{booktabs}
\usepackage[caption=false,font=normalsize,labelfont=sf,textfont=sf]{subfig}
\usepackage{textcomp}
\usepackage{stfloats}
\usepackage{url}
\usepackage{verbatim}
\usepackage{graphicx}
\usepackage{cite}
\usepackage{gensymb}

\begin{document}
\title{Secrecy Sum Rate Maximization for OIRS-Aided Visible Light Communications with Confidential Messages}

\author{\IEEEauthorblockN{Trinh K. Nguyen, Hung K. Hoang, Thanh V. Pham, \textit{Senior Member, IEEE} and Chuyen T. Nguyen}
\thanks{Trinh K. Nguyen, Hung K. Hoang, and Chuyen T. Nguyen are with Hanoi University of Science and Technology, Hanoi, Vietnam (e-mail: trinh2k5vd@gmail.com, hungtpls2003@gmail.com, chuyen.nguyenthanh@hust.edu.vn).}
\thanks{Thanh V. Pham is with Shizuoka University, Shizuoka, Japan (email: pham.van.thanh@shizuoka.ac.jp).}
}
\maketitle
\begin{abstract}
This paper investigates the secrecy sum-rate (SSR) performance of optical intelligent reflecting surface (OIRS)-assisted multi-user visible light communication (VLC) systems under line-of-sight (LoS) blockages. To mitigate physical obstructions and internal eavesdropping, a joint optimization problem is formulated to maximize the SSR through the co-design of the transmission precoder and OIRS units assignment. Due to the binary constraints and coupled variables, the problem is highly non-convex. To solve it efficiently, an alternating optimization (AO) framework integrating the concave-convex procedure (CCCP) and first-order Taylor approximations is developed. Simulation results demonstrate the convergence of the proposed algorithm and show that increasing the number of OIRS reflecting units yields significant SSR gains.
\end{abstract}


\section{Introduction}

Visible light communication (VLC) has emerged as a promising technology for next-generation indoor networks by exploiting the vast unlicensed optical spectrum and existing LED lighting infrastructure \cite{komine2004fundamental}. Besides supporting high-speed data transmission, VLC also offers strong potential for green communications and physical layer security (PLS) \cite{VLCpoop}. Nevertheless, VLC systems heavily rely on line-of-sight (LoS) propagation and are therefore highly vulnerable to physical blockages, which can severely attenuate or even interrupt communication links \cite{komine2004fundamental}. To address this issue, optical intelligent reflecting surfaces (OIRSs) have recently emerged as an effective solution for establishing controllable non-LoS (NLoS) links \cite{syl2023oirstuto}. By reconfiguring the optical propagation environment, OIRSs can bypass obstacles and maintain reliable connectivity in obstructed indoor scenarios.

Meanwhile, ensuring secure transmission in multi-user (MU) VLC systems remains a critical challenge. Although OIRSs improve connectivity, the additional reflection paths may also strengthen unintended wiretap channels, thereby increasing the risk of information leakage. As a result, properly configuring OIRS reflection states to enhance legitimate links while suppressing eavesdropping becomes a non-trivial problem.

Existing studies on OIRS-assisted VLC mainly focus on improving sum rate, spectral efficiency, and energy efficiency \cite{intro1},\cite{intro2}. Recent works addressing PLS consider only external eavesdroppers and neglect LoS blockages. Specifically, the study in \cite{intro3} maximizes the secrecy sum-rate (SSR) against a single external eavesdropper, whereas the work in \cite{intro4} investigates max-min secrecy in Non-Orthogonal Multiple Access (NOMA)-based VLC systems. However, jointly accounting for physical blockages and confidential message transmission in the presence of internal eavesdropping (i.e., eavesdropping among legitimate users) may fundamentally alter the optimization structure, rendering existing approaches unsuitable.

Motivated by these challenges, this letter investigates a secure MU-MISO VLC system that concurrently accounts for physical blockages and transmission of confidential messages. A joint optimization problem is formulated to maximize the confidential SSR through the co-design of transmission precoding and OIRS units configuration. However, the non-linear secrecy rate expressions involving both binary assignments and continuous precoding yield an intractable mixed-integer non-linear program (MINLP). To render this problem mathematically tractable, an efficient alternating optimization (AO) algorithm is developed, leveraging the concave-convex procedure (CCCP) and first-order Taylor approximations. Numerical simulations will be performed to demonstrate that the proposed scheme achieves robust secrecy performance under severe blockages while reliably converging to a high-quality suboptimal solution.

\emph{Notation}: Scalars, vectors, matrices, and tensors are represented by $x$, $\mathbf{x}$, $\mathbf{X}$, and $\mathbb{X}$, respectively. Moreover, $(\cdot)^{\text{T}}$, $\|\cdot\|_1$, $\| \cdot \|$ and $\text{diag}(\cdot)$ represent the transpose, $\ell_1$-norm, Euclidean norm, and diagonal operation, while $\langle \cdot, \cdot \rangle$ and $\odot$ denote the Frobenius inner product and the Hadamard product, respectively. Finally, $\mathbbm{1}_{\{\cdot\}}$ is the indicator function.

\section{System Model}

\begin{figure}[ht]
    \centering
\includegraphics[width=0.35\textwidth]{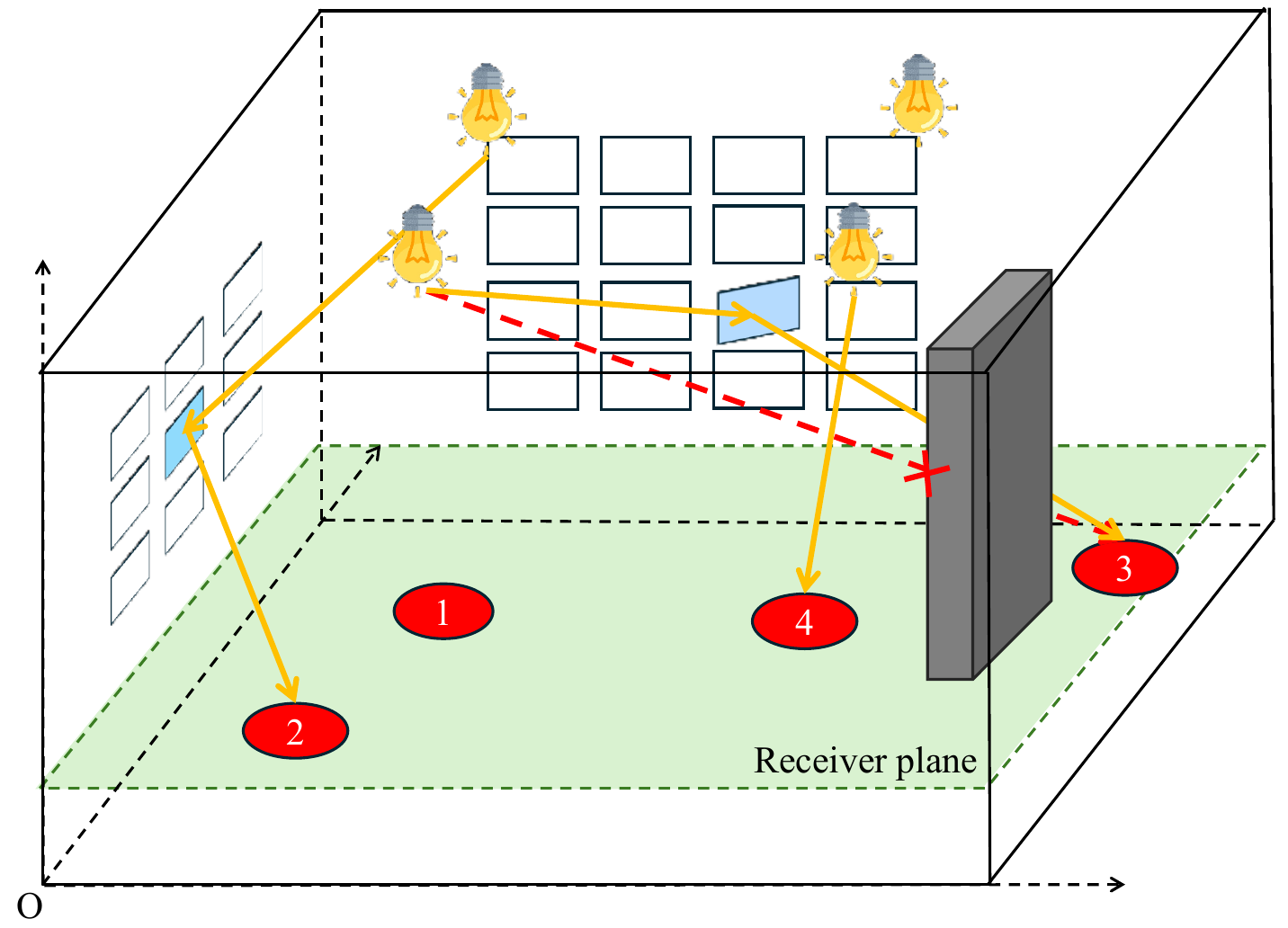}
    \caption{An example of the considered MU-MISO VLC system with $N_T = 4$, $K = 4$ and $M=25$ distributed on two walls.}
    \label{fig:1}
\end{figure}

The considered OIRS-aided MU-MISO VLC system, as shown in Fig.~\ref{fig:1}, consists of $N_T$ LED luminaries, $K$ independent users, and $M$ OIRS units. The LED luminaries are expected to transmit information to the users confidentially, meaning that each user is unable to decode any information intended for the others. Since the direct LoS paths between LEDs and users may be blocked by fixed obstacles, each user's photodiode (PD) is configured to receive both the direct LoS signals from the LEDs and the reflected NLoS signals from the assigned OIRS units.

\subsection{VLC System Description}

 Let $\mathbf{H}^{\text{LoS}} \in \mathbb{R}^{N_T \times K}$ denote the LoS channel matrix, whose $k$-th column is $\mathbf{h}^{\text{LoS}}_k=[h_{1,k},\;\ldots,\; h_{N_T,k}]^{\text{T}}\in \mathbb R^{N_T\times 1}$ represents the LoS channel vector between all $N_T$ LEDs and the $k$-th user. Here, $h_{n,k}^{\text{LoS}}$ denotes the LoS channel gain between the $n$-th LED luminary and the $k$-th user, which is given by \cite{komine2004fundamental} 
\begin{equation}        h_{n,k}^{\text{LoS}}=\frac{A_r}{d_{n,k}^2} L(\phi_{n,k}) T_s(\psi_{n,k})g(\psi_{n,k}) \cos(\psi_{n,k}),
\end{equation}
where $A_r$ is the active area of the PD, $d_{n,k}$ represents the distance between the $n$-th LED luminary and $k$-th user, $\phi_{n,k}$ and $\psi_{n,k}$ are the emission angle and the incident angle from $n$-th LED to the $k$-th user, respectively. Moreover, $L(\phi) = \frac{l+1}{2\pi}\cos^l(\phi)$ is the Lambertian emission intensity, where $l = -\frac{\ln(2)}{\ln(\Theta_{0.5})}$ denotes the Lambertian index, with $\Theta_{0.5}$ being the LED's semi-angle at half illuminance. Additionally, $T_s(\psi_{n,k})$ is the gain of optical filter, and $g(\psi_{n,k})$ denotes the gain of the optical concentrator, which is calculated as $g(\psi_{n,k}) = \mathbbm{1}_{\{0 \leq \psi_{n,k} \leq \Psi\}} \kappa^2/\sin^2(\Psi)$ \cite{komine2004fundamental}, where $\kappa$ is the refractive index of the concentrator and $\Psi$ denotes the optical field-of-view (FOV) of the PD.

Regarding the NLoS channel, let $\mathbb{H}^{\text{OIRS}} \in \mathbb{R}^{M \times N_T \times K}$ represents the complete OIRS reflection tensor. For the $k$-th user, the reflected channel matrix is defined as $\mathbf{H}_{k}^{\text{OIRS}} = [\mathbf{h}_{1,k}^{\text{OIRS}}, \ldots, \mathbf{h}_{N_T,k}^{\text{OIRS}}] \in \mathbb{R}^{M \times N_T}$, where $\mathbf{h}_{n,k}^{\text{OIRS}} = [h_{n,1,k}^{\text{OIRS}}, \ldots, h_{n,M,k}^{\text{OIRS}}]^{\text{T}} \in \mathbb{R}^{M \times 1}$ collects the link gains from the $n$-th LED via all $M$ OIRS units. Here, $h_{n,m,k}^{\text{OIRS}}$ represents the specific channel gain via the $m$-th OIRS unit, which is given as \cite{abdelhady2020visible}

\vspace{-0.3cm}
{\small
\begin{align}
h_{n,m,k}^{\text{OIRS}} &= 
     \frac{\delta A_r}{(d_{n,m} + d_{m,k})^2} L(\phi_{n,m}) T_s(\psi_{m,k}) \nonumber
    \\& \times g(\phi_{n,m},\psi_{m,k}) \cos(\psi_{m,k}),
\end{align}
}where $\delta$ denotes the reflectance coefficient of the OIRS unit; $d_{n,m}$ and $d_{m,k}$ are the distances between the $n$-th LED to the $m$-th OIRS unit and the $m$-th OIRS unit to the $k$-th user, respectively. $g(\phi_{n,m},\psi_{m,k})$ is expressed as
\begin{small}
\begin{equation}
    g(\phi_{n,m},\psi_{m,k}) = 
\begin{cases} 
\frac{\kappa ^2}{\sin^2(\Psi)}, & \hspace{-0.2cm} 0 \leq \psi_{m,k} \leq \Psi 
 \ \text{and} \ 0 \leq \phi_{n,m} \leq \Theta_{0.5},\\ 
0, & \hspace{-0.2cm} \psi_{m,k} > \Psi \ \text{or} \ \phi_{n,m} > \Theta_{0.5}.
\end{cases}
\end{equation}
\end{small}
It is noted that, under point-source illumination, each OIRS unit serves only one user at a time due to the law of specular reflection \cite{9756553}. To describe the LED-ORIS-user association, we introduce a binary tensor $\mathbb{F} \in \lbrace0,1\rbrace^{M\times N_T\times K}$ representing the association behavior of OIRS units and LED-PD pairs. Let $\mathbf{F}_k \in \{0, 1\}^{M \times N_T}$ denote the association matrix for the $k$-th user, where the entry $f_{n,m,k} = 1$ if the $m$-th OIRS unit reflects the signal from the $n$-th LED to the $k$-th user, and $0$ otherwise \cite{9526581}. Consequently, if $\mathbf{h}_k=[h_{1,k},\;\ldots,\;h_{N_T,k}]^{\text{T}}\in \mathbb{R}^{N_T\times 1}$ is denoted as the effective channel vector from LED arrays to the $k$-th user, it can be written as
\begin{equation}
\mathbf{h}_k=\mathbf h_k^{\text{LoS}}+\mathrm{diag}(\mathbf {H_k^{\text{OIRS}}}^{\text{T}}\mathbf F_k).
\end{equation}
\subsection{Signal Model}
Let $\mathbf{d} = [d_1,\;\ldots,\;d_K]^{\text{T}}\in \mathbb{R}^{K\times 1}$ be the transmitted data vector for all $K$ user, where each $M$-PAM symbols $d_k \in [-1, 1]$ modeled as zero-mean random variable with variance $\sigma_d^2$. The transmitted signal at the $n$-th LED is given by $s_n = \mathbf{v}_n \mathbf{d},$ where $\mathbf{v}_n = [w_{n,1}, \dots, w_{n,K}]$ represents the precoding vector for the $n$-th LED, and $\mathbf{w}_k = [w_{1,k}, \dots, w_{N_T,k}]^{\text{T}}$ is the precoder dedicated to the $k$-th user. For illumination, a DC bias $I_n^{\text{DC}}$ is added to $s_n$, forming the driving current $z_n = s_n + I_n^{\text{DC}}$. To maintain operation within the LED's linear dynamic range, $z_n$ must satisfy $0 \leq z_n \leq I_{\text{max}}$. Given that $|d_k| \leq 1$, the signal amplitude is bounded by $|s_n| \leq \|\mathbf{v}_n\|_1$. Consequently, to prevent clipping distortion while satisfying the linear range \cite{Elgala2009}, the precoder must satisfy 
\begin{equation}
\sum_{k=1}^K|w_{n,k}|\le\min(I_n^{DC},I_{max}-I_n^{DC})\label{precoder_constraint}.
\end{equation}

The emitted optical power of each LED luminary is given by $P_{n}^s= \eta (s_n +I_n^{DC})$, where $\eta$ as the LED conversion factor. At the receiver, by combining the optical signals from both the direct LoS path and the cascaded LED-OIRS-PD paths, the electrical signal received at the $k$-th user is given by
\begin{align}
        y_k = \gamma\eta(\mathbf h_k^{\text{T}}\sum_{i=1}^K\mathbf w_id_i +\mathbf h_k^{\text{T}}\mathbf I^{DC})+n_k ,
    \label{receive_signal}
\end{align}
where $\gamma$ is the PD responsivity and $\mathbf{I}^{DC}=[I_1^{DC},\;\ldots,\;I _{N_T}^{DC}]^{\text{T}}$. The receiver noise $n_k$ is a real-valued zero-mean Gaussian random process with variance defined in \cite{zeng2009}. To simplify the optimization problem in this work, we replace the noise variance with an upper bound $\hat{\sigma}_k^2$, thereby removing its dependence on $\mathbb{F}$. This upper bound is obtained using a fixed matrix $\hat{\mathbf{F}}_k$ determined through exhaustive search by selecting, for each LED, the OIRS units that maximize the reflected channel gain toward user $k$. Accordingly, $\hat{\sigma}_k^2$ is expressed as

\vspace{-0.35cm}
\begin{small}
\begin{align}
    \hat{\sigma}_k^2 &= 2\gamma e \eta\left(\mathbf h_k^{\text{LoS}}+\text{diag}({\mathbf{H}_k^{\text{OIRS}}}^{\text{T}}\hat{\mathbf{F}}_k)\right)^{\text{T}}\mathbf I^{DC} B \\&+ 4\pi e A_r \gamma \chi_{amb}(1-\cos(\Psi))B + i_{amp}^2 B, \nonumber
\end{align}
\end{small}where $e$, $B$, $\chi_{amb}$, $i_{amp}$ is the elementary charge, the modulation bandwidth, the ambient light photo-current and the pre-amplifier noise current density, respectively.

\subsection{Secrecy Sum Rate}
For demodulation, the DC component in the received signal in \eqref{receive_signal} is removed, yielding
\begin{equation}
    \bar{y}_k=\left[\mathbf h_k^{\text{LoS}}+\text{diag}(\mathbf {H_k^{\text{OIRS}}}^{\text{T}}\mathbf F_k)\right]^{\text{T}}\sum_{i=1}^K\mathbf w_id_i+\bar{n}_k,
\end{equation}
where $\bar{n}_k = \frac{n_k}{\eta \gamma}$.  A lower bound of achievable confidential secrecy rate of the $k$-th user is given as \cite{8472921}

\begin{small}
\begin{align}
    &R_{s,k} = \frac{1}{2}\log_2\hspace{-0.1cm}\left(\frac{1+\sum_{i=1}^Ka_k\left[(\mathbf h_k^{\text{LoS}}+\text{diag}({\mathbf H_k^{\text{OIRS}}}^{\text{T}}\mathbf F_k))^{\text{T}}\mathbf{w}_i\right]^2}{1+\sum_{i=1,i\ne k}^Kb_k\left[(\mathbf h_k^{\text{LoS}}+\text{diag}({\mathbf H_k^{\text{OIRS}}}^{\text{T}}\mathbf F_k))^{\text{T}}\mathbf{w}_i\right]^2}\right) \nonumber 
    \\&-\frac{1}{2}\log_2\left(1+\sum_{i=1,i\ne k}^Kb_i\left[(\mathbf h_i^{\text{LoS}}+\text{diag}({\mathbf H_i^{\text{OIRS}}}^{\text{T}}\mathbf F_i))^{\text{T}}\mathbf{w}_k\right]^2\right),
\end{align}
\end{small}where $a_k = \frac{\exp(2h_d)}{2 \pi \bar{\sigma}_k^2}$, $b_k=\frac{\sigma_d^2}{\bar{\sigma}_k^2}$, $\bar\sigma_k^2=\frac{\hat{\sigma}_k^2}{(\gamma\eta)^2}$  and $h_d$ being the differential entropy of $d$. Then, if $\mathbf W=[\mathbf w_1, \mathbf w_2,...,\mathbf w_K]\in R^{N_T\times K}$ is denoted as the precoder matrix, the secrecy sum rate (SSR) of the considered system is given by $\Phi(\mathbb{F},\mathbf{W})=\sum_{k=1}^K R_{s,k}(\mathbb{F},\mathbf{W})$.

\section{Secrecy Sum Rate Maximization}
We aim to maximize the overall SSR by jointly optimizing the OIRS assignment tensor $\mathbb{F}$ and the precoding matrix $\mathbf{W}$. In particular, this optimization problem is formulated as follows

\vspace{-0.3cm}
\begin{small}
\begin{subequations}\label{Problem_P0}
\begin{align}
\mathcal{P}\mathbf{0}:&\underset{\mathbb{F}, \mathbf{W}}{\text{maximize}}\quad\Phi(\mathbb{F},\mathbf{W}) ,
\\&\text{subject to} \quad \nonumber \\ &\qquad R_{s,k}(\mathbb{F},\mathbf{W}) \ge \lambda_k ,\label{P0_constraint_rate}
\\&\qquad\sum_{k=1}^K\mathbf{F_k}\mathbf{1}_{N_T} =\mathbf{1}_{M} \label{P0_constraint_crosslayer},
\\ &\qquad f_{n,m,k}=\lbrace0,1\rbrace ,\forall k \in K,\forall m \in M, \forall n \in N_T  \label{P0_constraint_binary},
\\ &\qquad\sum_{k=1}^K|w_{n,k}|\le\min(I_n^{DC},I_{max}-I_n^{DC})\label{P0_constraint_precoder},
\end{align}
\end{subequations}
\end{small}where $\mathbf{1}_{N_T}$ and $\mathbf{1}_M$ denote the all-ones column vectors of size $N_T$ and $M$, respectively.
Here, constraint \eqref{P0_constraint_rate} enforces the secrecy rate requirement, where $\lambda_k$ denotes the minimum threshold of the $k$-th user. Constraints \eqref{P0_constraint_crosslayer} and \eqref{P0_constraint_binary} govern the OIRS assignment process: the former ensures that each OIRS unit is assigned to only one LED-user pair to avoid SSR degradation, while the latter restricts the assignment variables to binary values. In addition, \eqref{P0_constraint_precoder} ensures that the transmitted signals remain within the LEDs' linear dynamic range to prevent clipping distortion. 

\vspace{-0.5cm}
\begin{table}[ht]
\begin{small}
\begin{subequations}
    \label{slack_problem}
    \begin{flalign} 
        \mathcal{P} \mathbf{1}:~ 
        \underset{\substack{{\mathbb{F},\mathbf{W}, \mathbf{r}, \mathbf{p}}}}{\text{maximize}} 
        \hspace{3mm} & \sum_{k=1}^{K} (r_{1,k} - r_{2,k} - r_{3,k}), 
        && \label{P1_objective}
    \end{flalign}      
    \vspace{-\baselineskip}
    \hspace{-0.5cm}
    \begin{minipage}{0.25\textwidth}
        \vspace{-0.65cm}
        \begin{align}
        & \nonumber\text{subject to}  \\
        & r_{1,k} \leq 0.5 \log_2(1 + p_{1,k}), \label{P1_cons_r1}
        \\& r_{2,k} \geq 0.5 \log_2(1 + p_{2,k}), \label{P1_cons_r2}
        \end{align}
    \end{minipage}
    \hfill
    \begin{minipage}{0.25\textwidth}
        \vspace{-0.3cm}
        \begin{align}
        &  r_{3,k} \geq 0.5 \log_2(1 + p_{3,k}), \label{P1_cons_r3}
        \\& r_{1,k} - r_{2,k} - r_{3,k} \geq \lambda_k, \label{P1_cons_r123}
        \end{align}
    \end{minipage}
    \vspace{0.2cm}
    \begin{flalign} 
        p_{1,k} \leq \sum_{i=1}^Ka_k \left( \left(\mathbf {h}_k^{\text{LoS}}+\text{diag}({\mathbf {H}_k^{\text{OIRS}}}^{\text{T}}\mathbf {F}_k)\right)^{\text{T}} \mathbf{w}_i\right)^2, && \label{P1_cons_p1}
    \end{flalign}  
    \vspace{-0.45cm}
    \begin{flalign}
        p_{2,k} \geq \sum_{i=1,i\ne k}^K b_k\left(\left(\mathbf h_k^{\text{LoS}}+\text{diag}({\mathbf H_k^{\text{OIRS}}}^{\text{T}}\mathbf F_k)\right)^{\text{T}}\mathbf{w}_i\right)^2, && \label{P1_cons_p2}
    \end{flalign}
    \vspace{-0.45cm}
    \begin{flalign}
        p_{3,k} \geq \sum_{i=1,i\ne k}^Kb_i\left(\left(\mathbf h_i^{\text{LoS}}+\text{diag}({\mathbf H_i^{\text{OIRS}}}^{\text{T}}\mathbf F_i)\right)^{\text{T}}\mathbf{w}_k\right)^2, && \label{P1_cons_p3}
    \end{flalign}
    \vspace{-0.45cm}
    \begin{flalign}
        \eqref{P0_constraint_crosslayer}, \eqref{P0_constraint_binary},\eqref{P0_constraint_precoder}. && \nonumber
    \end{flalign}
\end{subequations}
\end{small}
\vspace{-\baselineskip}
\end{table}

\vspace{-0.1cm}
It should be noted that the problem in \eqref{Problem_P0} is non-convex
due to the non-concave objective function and the non-convex constraints in \eqref{P0_constraint_rate}, \eqref{P0_constraint_binary}. To tackle this non-convexity, we introduce slack variables $\mathbf{r} = \{r_{1,k}, r_{2,k}, r_{3,k}\}$ and $\mathbf{p} = \{p_{1,k}, p_{2,k}, p_{3,k}\}$ to reformulate the problem to $\mathcal{P}\mathbf{1}$. Under this framework, the coupled subproblems associated with $\mathbb{F}$ and $\mathbf{W}$ are solved iteratively using alternating optimization (AO) and the concave-convex procedure (CCCP), as detailed in the following subsections.

\subsection{OIRS assignment optimization}
In this part, the OIRS assignment tensor $\mathbb{F}$ will be optimized for a given precoder matrix $\mathbf{W}$, which can be expressed as
\vspace{-0.1cm}
\begin{align}
\mathcal{P}\mathbf{2a}:&\underset{\mathbb{F},\mathbf{r},\mathbf{p}}{\text{maximize}} \quad \eqref{P1_objective}, \nonumber
\\&\text{subject to} \quad \eqref{P1_cons_r1}-\eqref{P1_cons_p3},\eqref{P0_constraint_crosslayer}, \eqref{P0_constraint_binary}. \nonumber
\end{align}
The discrete nature of \eqref{P0_constraint_binary} renders the optimization problem NP-hard due to its mixed-integer programming (MIP) structure. To address this challenge, we adopt the penalty-based approach proposed in \cite{9525467}. Specifically, constraint \eqref{P0_constraint_binary} is first relaxed into continuous bounds within $[0,1]$, after which a penalty term is incorporated into the objective function encourage convergence toward binary solutions. Accordingly, the problem can be reformulated as

\vspace{-0.3cm}
\begin{small}
\begin{subequations} \label{P2b}
\begin{align}
\mathcal{P}\mathbf{2b}:&\underset{\mathbb{F},\mathbf{r},\mathbf{p}}{\text{maximize}} \quad \sum_{k=1}^{K} (r_{1,k} -r_{2,k} - r_{3,k})\label{P2a_penalty_term} \\&-\nonumber\lambda\sum_{k=1}^{K}\sum_{m=1}^{M}\sum_{n=1}^{N_T}(f_{n,m,k}-f_{n,m,k}^2),
\\&\text{subject to} \quad \eqref{P1_cons_r1}-\eqref{P1_cons_p3},\eqref{P0_constraint_crosslayer}, \nonumber
\end{align}
\end{subequations}
\end{small}where $\lambda$ is the sufficiently large coefficient. After reformulation, constraints \eqref{P1_cons_r2}, \eqref{P1_cons_r3}, \eqref{P1_cons_p1} and the objective function \eqref{P2a_penalty_term} remain non-convex. To address this issue, first-order Taylor approximations are applied at the $t$-th iteration to linearize both the penalty term and the remaining non-convex constraints as follows

\vspace{-0.55cm}
\begin{small}
\begin{subequations} \label{P2c}
\begin{align}
&\mathcal{P}\mathbf{2c}: \underset{\mathbb{F},\mathbf{r},\mathbf{p}}{\text{maximize}} \quad \sum_{k=1}^{K} (r_{1,k} - r_{2,k} - r_{3,k})-\lambda\sum_{k=1}^{K}\sum_{m=1}^{M}\sum_{n=1}^{N_T} \nonumber \\&\left[(f_{n,m,k}^{(t-1)}-f_{n,m,k}^{2(t-1)})+(1-2f_{n,m,k}^{(t-1)})(f_{n,m,k}^{(t)}-f_{n,m,k}^{(t-1)})\right],
\\& \text{subject to} \nonumber 
\\& p_{1,k}(\mathbf F_k)  \le p_{1,k}(\mathbf F_k^{(t-1)})+ \left\langle \nabla p_{1,k}(\mathbf F_k^{(t-1)}),\mathbf F_k -\mathbf F_k^{(t-1)} \right\rangle 
\\&r_{s,k} \ge 0.5 \log_2 \left( 1 + p_{s,k}^{(t-1)} \right)+ \frac{ p_{s,k}^{(t)} - p_{s,k}^{(t-1)} }{ 2 \ln(2) \left( 1 + p_{s,k}^{(t-1)} \right) }, s\in  \{2,3\}
\\&\eqref{P0_constraint_crosslayer}, \eqref{P1_cons_r1}, \eqref{P1_cons_r123}, \eqref{P1_cons_p2}, \eqref{P1_cons_p3}, \nonumber
\end{align}
\end{subequations}
\end{small}where $\nabla p_{1,k}(\mathbf F_k^{(t-1)})=\sum_{i=1}^K2a_k(\text{Tr}(\mathbf F_k^{(t-1)}({\mathbf H_k^{\text{OIRS}}}^{\text{T}}\odot(\mathbf w_i1_M^{\text{T}})))+{\mathbf h_k^{\text{LoS}}}^{\text{T}} \mathbf w_i)({\mathbf H_k^{\text{OIRS}}}^{\text{T}}\odot(\mathbf w_i 1_M^{\text{T}}))$. Problem $\mathcal{P}\mathbf{2c}$ is convex, thus can be solved using \texttt{CVX}. Accordingly, the problem $\mathcal{P}\mathbf{2b}$ is solved iteratively via the CCCP framework by successively updating $\mathcal{P}\mathbf{2c}$ until convergence as presented in \textbf{Algorithm \ref{alg:CCCP1}}.

\begin{small}
\begin{algorithm}[t]
\caption{CCCP-type algorithm for solving $\mathcal{P}\mathbf{2b}$}
\label{alg:CCCP1}
\begin{algorithmic}[1]
\State \textbf{Choose} the maximum number of iteration $L_{\max,1}$ and the error tolerance $\varepsilon_1 > 0$.
\State \textbf{Choose} feasible initial points $\mathbb{F}^{(0)}$, $p_{2,k}^{(0)}$, $p_{3,k}^{(0)}$ to $\mathcal{P}\mathbf{2c}$.
\State \textbf{Set} $t = 1$.
\While{convergence == \textbf{False} \textbf{and} $t \leq L_{\max,1}$}
    \State \textbf{Solve} $\mathcal{P}\mathbf{2c}$ for $\mathbb{F}^{(t)}$, $p_{2,k}^{(t)}$, $p_{3,k}^{(t)}$ using $\mathbb{F}^{(t-1)}$, $p_{2,k}^{(t-1)}$, $p_{3,k}^{(t-1)}$ obtained from the previous iteration.
    \If{$\frac{\|\mathbb{F}^{(t)} - \mathbb{F}^{(t-1)}\|}{\|\mathbb{F}^{(t)}\|} \leq \varepsilon_1$ \textbf{and} $\frac{\|\mathbf{p}_{s,k}^{(t)} - \mathbf{p}_{s,k}^{(t-1)}\|}{\|\mathbf{p}_{s,k}^{(t)}\|} \leq \varepsilon_1$, for $s\in\lbrace2,3\rbrace$}
        \State convergence = \textbf{True}
        \State $\mathbb{F}^* = \mathbb{F}^{(t)}$
        \State $p_{s,k}^{*} = p_{s,k}^{(t)}$, for $s\in \{2,3\}$
    \Else
        \State convergence = \textbf{False}
    \EndIf
    \State $t := t + 1$
\EndWhile
\State \textbf{return} the optimal value $\mathbb{F}^*$.
\end{algorithmic}
\end{algorithm}
\vspace{-\baselineskip}
\end{small}

\subsection{Precoder optimization}
After obtaining the optimized OIRS assignment tensor $\mathbb{F}^{*}$, the precoding matrix $\mathbf{W}$ is updated. The corresponding optimization subproblem is formulated as

\vspace{-0.4cm}
\begin{small}
\begin{align}
\mathcal{P}\mathbf{3a}:& \underset{\mathbf{W},\mathbf{r},\mathbf{p}}{\text{maximize}} \quad \eqref{P1_objective} \nonumber \\
& \text{subject to} \quad \eqref{P1_cons_r1}-\eqref{P1_cons_p3},\eqref{P0_constraint_precoder} .\nonumber
\end{align}
\end{small}
$\mathcal{P}\mathbf{3a}$ is observed to be non-convex due to the non-convex constraints \eqref{P1_cons_r2}, \eqref{P1_cons_r3}, \eqref{P1_cons_p1}. A similar first-order Taylor approximation is applied at the $t$-th iteration, resulting in

{\small
\vspace{-0.4cm}
\begin{subequations}\label{P3b}
\begin{align}
&\mathcal{P}\mathbf{3b}: \underset{\mathbf{W},\mathbf{r},\mathbf{p}}{\text{maximize}} \quad \eqref{P1_objective}, \nonumber \\
& \text{subject to} \nonumber \\
& p_{1,k} \leq \sum_{i=1}^{K} a_k \Big( (\mathbf{h}_k^{\text{T}} \mathbf{w}_i^{(t-1)})^2 + 2 (\mathbf{w}_i^{(t-1)})^\mathrm{T} \mathbf{h}_k \mathbf{h}_k^\mathrm{T} (\mathbf{w}_i^{(t)} - \mathbf{w}_i^{(t-1)}) \Big), \label{P3_constraint_Taylor_p1} \\
& r_{s,k} \geq 0.5 \log_2 (1 + p_{s,k}^{(t-1)}) + \frac{p_{s,k}^{(t)} - p_{2,k}^{(t-1)}}{2 \ln(2) (1 + p_{s,k}^{(t-1)})},s\in\lbrace2,3\rbrace\label{P3_constraint_Taylor_r23} \\
&\eqref{P0_constraint_precoder}, \eqref{P1_cons_r1}, \eqref{P1_cons_r123},\eqref{P1_cons_p2}, \eqref{P1_cons_p3},  \nonumber
\end{align}
\end{subequations}}where $\mathbf{w}_i^{(t-1)},p_{2,k}^{(t-1)}$ and $p_{3,k}^{(t-1)}$ are the solutions obtained from the previous iteration. Since $\mathcal{P}\mathbf{3b}$ is convex, it can be efficiently solved using \texttt{CVX}. Consequently, the original subproblem $\mathcal{P}\mathbf{3a}$ is addressed by iteratively updating and solving $\mathcal{P}\mathbf{3b}$ within the CCCP framework until convergence, as summarized in \textbf{Algorithm~\ref{alg:CCCP2}}. The obtained $\mathbf{W}^*$ is then substituted back into the first subproblem, and the process is repeated until $\mathcal{P}\mathbf{1}$ converges. The overall iterative procedure is summarized in \textbf{Algorithm \ref{AO}}.

\vspace{-0.2cm}
\begin{small}
\begin{algorithm}[t]
\caption{CCCP-type algorithm for solving 
$\mathcal{P}\mathbf{3b}$}
\label{alg:CCCP2}
\begin{algorithmic}[1]
\State \textbf{Choose} the maximum number of iteration $L_{\max,2}$ and the error tolerance $\varepsilon_2 > 0$.
\State \textbf{Choose} feasible initial points $\mathbf{W}^{(0)}$, $p_{2,k}^{(0)}$, $p_{3,k}^{(0)}$ to $\mathcal{P}\mathbf{3b}$.
\State \textbf{Set} $t = 1$.
\While{convergence == \textbf{False} \textbf{and} $t \leq L_{\max,2}$}
    \State \textbf{Solve} $\mathcal{P}\mathbf{3b}$ for $\mathbf{W}^{(t)}$, $p_{2,k}^{(t)}$, $p_{3,k}^{(t)}$ using $\mathbf{W}^{(t-1)}$, $p_{2,k}^{(t-1)}$, $p_{3,k}^{(t-1)}$ obtained from the previous iteration.
    \If{$\frac{\|\mathbf{W}^{(t)} - \mathbf{W}^{(t-1)}\|}{\|\mathbf{W}^{(t)}\|} \leq \varepsilon_2$ \textbf{and} $\frac{\|\mathbf{p}_{s,k}^{(t)} - \mathbf{p}_{s,k}^{(t-1)}\|}{\|\mathbf{p}_{s,k}^{(t)}\|} \leq \varepsilon_2$, for $s\in\lbrace2,3\rbrace$}
        \State convergence = \textbf{True}
        \State $\mathbf{W}^* = \mathbf{W}^{(t)}$
        \State $p_{s,k}^*=p_{s,k}^{(t)}$, for $s\in \{2,3\}$
    \Else
        \State convergence = \textbf{False}
    \EndIf
    \State $t := t + 1$
\EndWhile
\State \textbf{return} the optimal value $\mathbf{W}^*$.
\end{algorithmic}
\end{algorithm}
\end{small}

\begin{small}
\begin{algorithm}[ht]
\caption{AO Algorithm for solving $\mathcal{P}\mathbf{0}$}
\label{AO}
\begin{algorithmic}[1]
\State \textbf{Choose} maximum iterations $L_{\max,3}$ and the error tolerance $\varepsilon_3 > 0$.
\State \textbf{Set} $t = 1$.
\Repeat
    \State Given $\mathbf{W}^{(t-1)}$, using \textbf{Algorithm \ref{alg:CCCP1}} to obtain $\mathbb{F}^{(t)}$ .
    \State Given $\mathbb{F}^{(t-1)}$, using \textbf{Algorithm \ref{alg:CCCP2}} to obtain $\mathbf{W}^{(t)}$.
    \State Calculate $\Phi(\mathbb{F}^{(t)},\mathbf{W}^{(t)})$ based on $\mathbf{W}^{(t)}$ and $\mathbb{F}^{(t)}$.
    \State Compute $\small{\epsilon_r^{(t)} = \frac{|\Phi(\mathbb{F}^{(t)},\mathbf{W}^{(t)}) - \Phi(\mathbb{F}^{(t-1)},\mathbf{W}^{(t-1)})|}{|\Phi(\mathbb{F}^{(t)},\mathbf{W}^{(t)})|}}$.
    \State $t := t + 1$
\Until{$\small{\epsilon_r^{(t)}} \leq \varepsilon_3$ \textbf{or} $t == L_{\max,3}$}
\State \textbf{Set} $\mathbf{W}^* = \mathbf{W}^{(t)}$, $\mathbb{F}^* = \mathbb{F}^{(t)}$.
\State \textbf{return} the optimal values $\mathbf{W}^*, \mathbb{F}^*$.
\end{algorithmic}
\end{algorithm}
\end{small}

\section{Simulation Results and Discussions}
In this section, the convergence behaviors and the achievable sum secrecy rate of the proposed solution are evaluated. The simulation considers a $5\text{m long} \times 5\text{m width} \times 3\text{m height}$ indoor environment, where all users are distributed on a horizontal plane at a height of $0.5 \text{m}$ from the floor. Four LEDs are deployed on the ceiling at coordinates $(1.5,~3.5,~3),~(1.5,~1.5,~3),~(3.5,~1.5,~3)$ and $(3.5,~3.5,~3)$. To model the LoS blockage, an obstacle of size $1\text{m long} \times 1\text{m width} \times 2.5\text{m height}$ is placed within the boundaries $3.5 \le x \le 4.5\text{ m}$, $3.0 \le y \le 4.0\text{ m}$, and $0 \le z \le 2.5\text{ m}$. Furthermore, two OIRSs are deployed as uniform planar arrays with a $0.2\text{ m}$ unit spacing, centered at $(2.5, 5.0, 1.5)$ on the $y=5$ plane and $(0.0, 2.5, 1.5)$ on the $x=0$ plane, respectively. Unless otherwise stated, the remaining simulation parameters follow \cite{son2021}.

\begin{figure}[ht]
    \centering
    \includegraphics[width=0.4\textwidth]{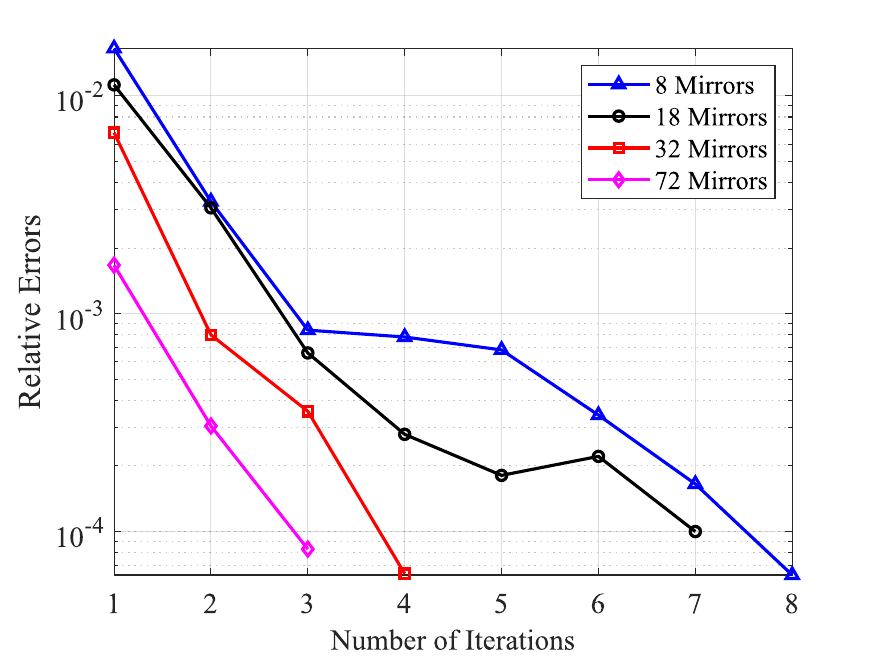}
    \caption{Relative error with respect to the number of iterations.}
    \label{fig:Relative_error}
\end{figure}
Figure~\ref{fig:Relative_error} illustrates the convergence behavior of the proposed AO-based scheme under different OIRS unit configurations. To evaluate the convergence behavior of the proposed scheme, the relative error of the objective value in $\mathcal{P}0$ is employed, which is defined as $\small{\epsilon_r^{(t)} = \frac{|\Phi(\mathbb{F}^{(t)},\mathbf{W}^{(t)}) - \Phi(\mathbb{F}^{(t-1)},\mathbf{W}^{(t-1)})|}{|\Phi(\mathbb{F}^{(t)},\mathbf{W}^{(t)})|}}$. As observed in all cases, $\small{\epsilon_r^{(t)}}$ decreases monotonically from approximately $10^{-2}$ to below the convergence threshold of $10^{-4}$, confirming the numerical stability of the algorithm. The convergence speed is strongly affected by the OIRS scale: the configuration with $72$ units converges within only $3$ iterations, whereas the $8$-units case requires up to $8$ iterations. These results demonstrate the reliable convergence of the proposed framework to a high-quality suboptimal solution.

\begin{figure}[ht]
    \centering
\includegraphics[width=0.4\textwidth]{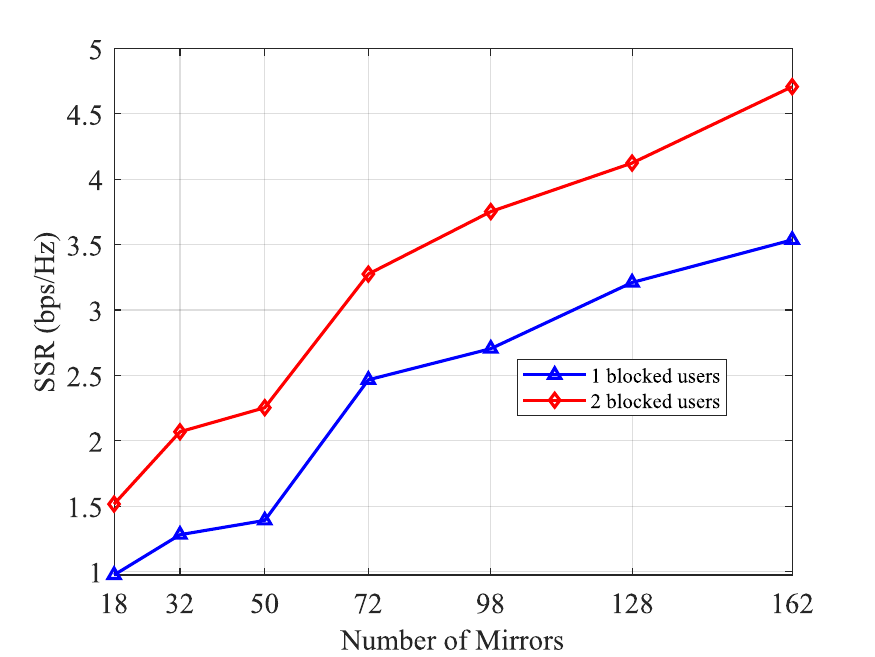}
    \caption{SSR versus the number of OIRS units.}
    \label{SSRvsOIRS}
\end{figure}
Figure~\ref{SSRvsOIRS} illustrates the confidential SSR versus the number of OIRS units, for the cases of one and two blocked users. In both scenarios, the SSR increases monotonically with the {number of OIRS units}, confirming that larger OIRSs can collect and redirect more optical power toward the intended users. Specifically, as the number of OIRS units increases from $18$ to $160$, the SSR rises from approximately $1$ to $3.6$ bps/Hz in the one-blocked-user case, and from $1.5$ to $4.7$ bps/Hz in the two-blocked-user case. Moreover, the two-blocked-user scenario is observed to consistently achieves a higher SSR than the one-blocked-user case in this figure. This behavior can be attributed to the simplified reflection model adopted in this work, where OIRS reflections are assumed to form highly directional specular beams toward the intended users, thereby reducing inter-user information leakage compared with wide-coverage LoS links. Under this assumption, increasing the number of users relying on reflected OIRS links tends to improve secrecy performance. Nevertheless, the actual optical propagation and reflection characteristics of OIRS-assisted VLC systems are considerably more complex in practice. A more accurate characterization of these physical effects and their impact on secrecy performance will be investigated in future work.
\section{Conclusion}
In this letter, we investigated an OIRS-assisted secure multi-user MISO VLC system under LoS blockages and internal eavesdropping threats. To maximize the SSR, a joint optimization framework was developed for the transmission precoder and OIRS units assignment. The resulting highly non-convex problem was addressed using an AO-based framework, where the decoupled subproblems were efficiently solved via the CCCP algorithm. Numerical results demonstrated that the proposed scheme effectively enhances multi-user secrecy performance while mitigating the impact of physical blockages. Furthermore, increasing the {number of OIRS units} was shown to provide substantial SSR gains, highlighting the potential of OIRS-assisted VLC for robust physical-layer security in indoor wireless networks.
\bibliographystyle{IEEEtran}
\bibliography{references}
\end{document}